\documentclass[%
 reprint,
superscriptaddress,
 amsmath,amssymb,
 aps,
]{revtex4-2}

\usepackage{dcolumn}
\usepackage{bm}

\usepackage[utf8]{inputenc}
\usepackage[T1]{fontenc}
\usepackage[version=4]{mhchem}
\usepackage{graphicx}
\usepackage{placeins}
\usepackage[usenames, dvipsnames]{xcolor}
\usepackage{url}
\usepackage{hyperref}
\usepackage{ulem}
\usepackage{comment}

\begin{document}

\title{Structural Descriptors and Information Extraction from X-ray Emission Spectra: Aqueous Sulfuric Acid}

\author{E. A. Eronen}
\affiliation{University of Turku, Department of Physics and Astronomy, FI-20014 Turun yliopisto, Finland}%
\author{A. Vladyka}%
\affiliation{University of Turku, Department of Physics and Astronomy, FI-20014 Turun yliopisto, Finland}%
\author{Ch. J. Sahle}%
\affiliation{ESRF, The European Synchrotron, 71 Avenue des Martyrs, CS40220, 38043 Grenoble Cedex 9, France}%
\author{J. Niskanen}
\affiliation{University of Turku, Department of Physics and Astronomy, FI-20014 Turun yliopisto, Finland}%

\date{\today}

\begin{abstract}
Machine learning can reveal new insights into X-ray spectroscopy of liquids when the local atomistic environment is presented to the model in a suitable way. Many unique structural descriptor families have been developed for this purpose. We benchmark the performance of six different descriptor families using a computational data set of 24200 sulfur K$\beta$ X-ray emission spectra of aqueous sulfuric acid simulated at six different concentrations. We train a feed-forward neural network to predict the spectra from the corresponding descriptor vectors and find that the local many-body tensor representation, smooth overlap of atomic positions and atom-centered symmetry functions excel in this comparison. We found a similar hierarchy when applying the emulator-based component analysis to identify and separate the spectrally relevant structural characteristics from the irrelevant ones. In this case, the spectra were dominantly dependent on the concentration of the system, whereas adding the second most significant degree of freedom in the decomposition allowed for distinction of the protonation state of the acid molecule.

\end{abstract}

\maketitle


\section{Introduction}
The liquid phase allows for the movement of solvent and solute molecules while simultaneously having strong interactions among them. This leads to a distribution of possible local structures, and respective local electronic Hamiltonians. Computations have shown that these local environments yield significantly different X-ray spectra, while their ensemble mean is needed for a match with the corresponding experiment \cite{Wernet2004, Leetmaa2010, Ottosson2011a, Sahle2013, Niskanen2016, Niskanen2017, Niskanen2019, VazdaCruz2019}. Thus the changes in the experimental spectrum can be connected to corresponding changes in the local structural distribution. However, the complexity of this problem calls for sophisticated methods able to distinguish between relevant and irrelevant information. Recent developments in computational resources and machine learning (ML) have opened new paths for these investigations \cite{Timoshenko2017, Timoshenko2019, Liu2019, Andrejevic2022,Niskanen2022a, Niskanen2022, Kwon2023, Vladyka2023, Eronen2024}.

Because raw atomic coordinates $\mathbf{R}$ are unsuitable for contemporary ML, numerous families of descriptors $\mathbf{D}(\mathbf{R})$ have been developed to encode the structural information into a useful input \cite{Behler2016, Himanen2020, Huo2022, Behler2011, Bartok2013, Rupp2012, Hansen2015,Thompson2015,Wood2018, Drautz2019, Chandrasekaran2019, Langer2022,Khan2023,Middleton2024}. While each of these representations might perform well for some tasks, they are not necessarily equally fit for every situation. Beside the ML performance, interpretability of the descriptor is a key consideration for studies of actual structural information content of, {\it e.g.} X-ray spectra \cite{Eronen2024}. The study of the spectrum-to-structure inverse problem is indeed highly dependent on the descriptor, which needs to include physically relevant features that are meaningful not only to ML models but also to human researchers.

Spectrum prediction by ML, {\it i.e.} finding a suitable function for spectrum $\mathbf{S}(\mathbf{D}(\mathbf{R}))$, is a more viable task than structure prediction (finding function $\mathbf{R}(\mathbf{S})$) \cite{Niskanen2022a}, possibly because the former is not an injective function. Moreover, some structural characteristics of the system are spectrally irrelevant. While the behavior of a spectrum can be captured by a well-performing feed-forward neural network (NN), the knowledge remains hidden in the respective weight matrices and bias vectors. To this end, the NN is useful only for predicting the outcome for new input, {\it i.e.} for emulation. Emulator-based component analysis (ECA) \cite{Niskanen2022} is an approach to extract knowledge contained by a ML model and a given data set. With the help of a fast emulator such as an NN, the method iteratively finds the structural dimensionality reduction for maximal explained spectral variance. The resulting basis vectors provide an exhaustive input feature selection, which not only points out spectrally relevant ones but also their collaborative effect \cite{Eronen2024}. Moreover, approximate structural reconstruction from spectra can be done by first reconstructing the few spectrally dominant latent coordinates, and then taking an expansion with the respective basis vectors \cite{Vladyka2023}. Our previous work on N K-edge X-ray absorption spectroscopy of aqueous triglycine \cite{Eronen2024} showed that ECA greatly outperforms principal component analysis (PCA) \cite{Jolliffe2016} of structural data in terms of covered spectral variance. Furthermore, a recent study using encoder-decoder neural networks supports the validity of ECA \cite{Passilahti2024}. While the advantage of this method is its lack of need for any prior hypothesis, it gives rise to another requirement for a descriptor: decomposability into a few dominant contributions \cite{Eronen2024}.

We explore structural information content of simulated sulfur K$\beta$ X-ray emission spectra (XES) of aqueous sulfuric acid. To create the data set we sample atomistic local structures from {\it ab initio} molecular dynamics (AIMD) simulations at six different concentrations and calculate spectra for these structures. We first assess a total of six structural descriptor families in terms of spectrum prediction performance by an NN. For a fair comparison between the descriptor families, we allocate equal computational resources to the joint hyperparameter--NN architecture search for each of them. Next, we identify the spectrally dominant structural degrees of freedom using ECA, and study the performance of the best descriptor of each family in the task. A sulfuric acid molecule can exist in one of three protonation states, which we find to be distinguishable from the XES after rank-two decomposition, in which the first rank covers intermolecular interaction given by the concentration. Our results highlight the need for identification of relevant structural degrees of freedom for reliable interpretation of X-ray spectra. Moreover, they raise a call for methods of obtaining simple structural information from contemporary structural descriptors, that may have a notably abstract mathematical form.

\section{Methods}

\begin{table*}[]
\centering
\caption{Details of the AIMD simulations for each number concentration denoted as the number of acid molecules versus the number of water molecules: the molarity, simulation box length, production run duration, sampling interval for the snapshots, the total number of snapshots N$_{\mathrm{snapshots}}$, the total number of emission sites for which the X-ray emission spectra are calculated N$_{\mathrm{spectra}}$, and the fractional abundancies of protonation states 0 (SO$_4^{2-}$), 1 (HSO$_4^{1-}$) and 2 (H$_2$SO$_4$). A few structures had the protonation state of 3 which is omitted from the table.\label{tab:simdetails}}
\begin{tabular}{lccccccccc}
Num. conc. & Molarity [M] & Box L [\AA] & Duration [ps] & Sampling [fs] & N$_{\mathrm{snapshots}}$ & N$_{\mathrm{spectra}}$  & SO$_4^{2-}$ [\%]& HSO$_4^{1-}$ [\%]& H$_2$SO$_4$ [\%]\\
\hline
1v63  & 0.9   & 12.50  & 100 & 62.5 & 1600 & 1600  & 96.2 & 3.8 & 0.0 \\ 
6v54  & 4.9   &  12.66 & 50 & 62.5 & 800 & 4800  & 71.4 & 28.5 & 0.1 \\
12v36 & 10.1  &  12.55 & 50 & 125 & 400 & 4800  & 17.2 & 79.3 & 3.5 \\
20v20 & 15.3  &  12.95 & 50 & 250 & 200 & 4000  & 2.0 & 74.8 & 23.1 \\
21v7  & 17.5  &  12.58 & 50 & 250 & 200 & 4200  & 0.0 & 28.7 & 70.8 \\
24v0  & 18.6  & 12.90 & 50 & 250 & 200 & 4800  & 0.0 & 0.0 & 99.6 \\
\hline
Total & & & & & & 24200\\
\hline
\hline
\end{tabular}
\end{table*}

We base the study on structure--spectrum data pairs obtained from AIMD simulations and subsequent spectrum calculations resulting in 24200 data points. We encode the structural data using six different descriptor families and study the resulting performance of ML and subsequent ECA.

\subsection{Simulations}

We extended the AIMD runs of Ref.~\citenum{Niskanen2015} for structural sampling from six concentrations. The details of the AIMD runs are presented in Table~\ref{tab:simdetails}. The simulations for the NVT ensemble were run using the CP2K software \cite{Kuhne2020} and Kohn-Sham density functional theory (DFT) with Perdew--Burke--Ernzerhof (PBE) exchange correlation potential \cite{pbe}. The AIMD runs utilized periodic boundary conditions, Goedecker--Teter-Hutter (GTH) pseudopotentials \cite{Goedecker1996,Hartwigsen1998,Krack2005} and triple-$\xi$ TZVP-GTH basis set delivered with the software.

We computed the XES for every sulfur site of each sampled snapshot using the projector-augmented-wave (PAW) method \cite{Enkovaara2010} with plane wave basis and density functional theory (DFT) implemented in GPAW version 22.1.0 \cite{Mortensen2005,Enkovaara2010,Larsen2017}. We used periodic boundary conditions, the PBE exchange correlation potential and a 600~eV plane wave energy cutoff (for justification, see Supplementary Information). 

The spectrum calculations applied transition potential DFT \cite{Triguero1998} in a fashion motivated in Ref. \citenum{Ljungberg2011}. First, we computed the neutral ground state of each snapshot and emission lines for each site on a relative energy scale. We then calibrated the individual spectra on the absolute energy scale by a $\Delta$-DFT calculation for the highest transition. This procedure builds on calculation with one valence vacancy and on respective calculations for the full core hole at each site, repeated for each snapshot. We convoluted the obtained energy--intensity pairs (a stick spectrum) with a Gaussian of full width at half maximum of 1.5~eV and then presented the spectra on a grid with bin width of 0.075~eV.

\subsection{Data preprocessing}

\begin{figure*}[]
    \centering
    \includegraphics[width=\textwidth]{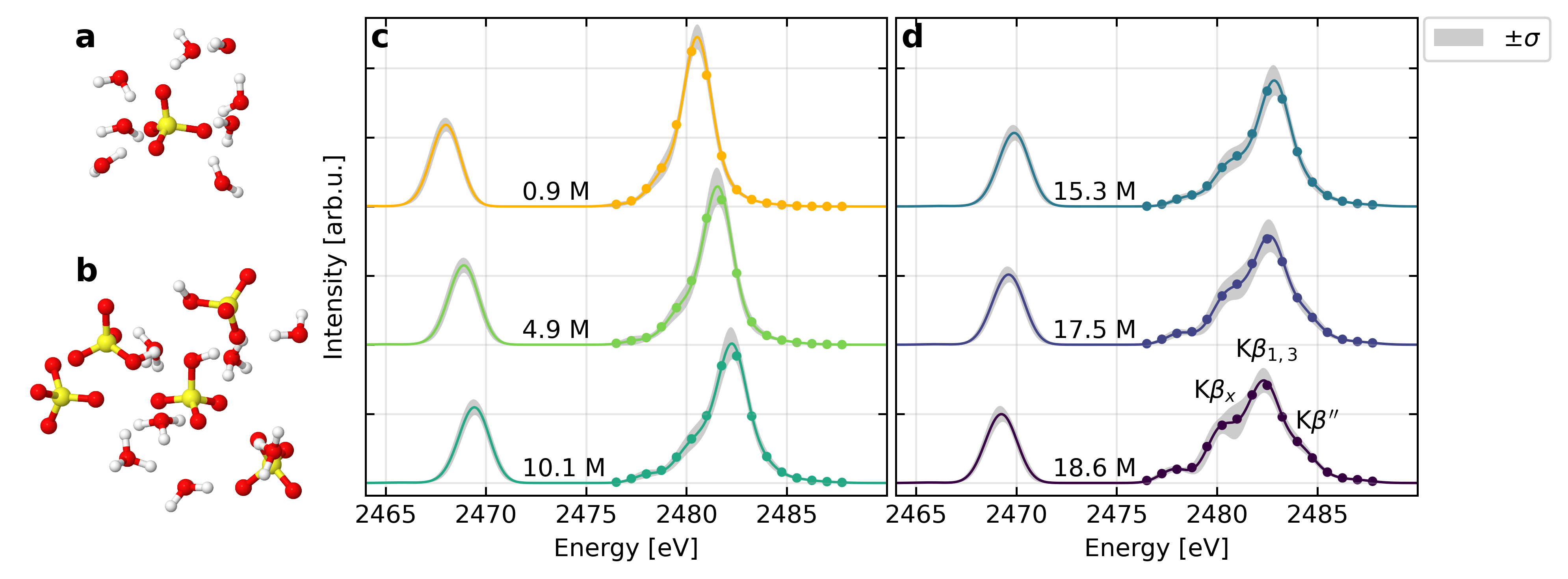}
    \caption{Simulation results of aqueous sulfuric acid. {\bf a,\,b:} Two sample structures for 0.9\,M and 15.3\,M, respectively, prepared with the Jmol software \cite{jmol}. Only molecules within 3\,\AA{} distance from the central molecule are shown for clarity. {\bf c,\,d:} Computational ensemble mean X-ray emission spectrum for each concentration with the standard deviation $\sigma$ shown as grey shaded area. As an example we show the location of the features K$\beta_\mathrm{x}$, K$\beta_\mathrm{1,3}$ and K$\beta''$ \cite{Mori2010} for the highest concentration spectrum in panel {\bf d}. In addition, the coarsened grid points used for the target spectra of the machine-learning-based analysis are shown on the mean spectra.}
    \label{fig:spectra}
\end{figure*}

For ML and further analysis, we took the portion with notable spectral intensity around the peak group K$\beta_\mathrm{x}$, K$\beta_\mathrm{1,3}$, and K$\beta''$ (see Figure \ref{fig:spectra} and {\it e.g.} Ref.~\citenum{Mori2010}), and coarsened it by integration to a new grid with bin width of 0.75~eV leading to target spectra $\mathbf{S}$ represented as vectors of 16 components. We chose this grid to be as coarse as possible while still containing the relevant spectral features due to three reasons: (i) the fewer output values simplify the ML problem; (ii) individual data points are more independent without significant loss of information; (iii) we hope to avoid over-interpretation through overly detailed analysis of simulations \cite{Niskanen2022a}. The last point follows the principle of correspondence {\it i.e.} analyzing simulations only to a degree in which they reproduce the experiment.

We evaluated the protonation state of each acid molecule as in Ref.~\citenum{Niskanen2015}. An oxygen atom was considered to belong to the molecule if it was at most $2.0~\text{\AA}$ from the respective sulfur atom. A hydrogen atom, in turn, was considered a part of the acid molecule if it was closer than $1.3~\text{\AA}$ from any of its oxygen atoms. Throughout the simulations, the acid molecules always had four oxygen atoms with the distribution of forms: 5878 \ce{SO4^2-}, 9435 \ce{HSO4^-}, 8847 \ce{H2SO4}, and 40 \ce{H3SO4^+} corresponding to protonations states from 0 to 3, respectively.

In this work we study six different descriptor families for encoding the local atomistic structure $\mathbf{R}$ around the emission site S$_\mathrm{em}$ into a vector of features $\mathbf{D}(\mathbf{R})$. We used the implementation within the DScribe-package (version 2.0.1) \cite{Himanen2020,dscribe2} for the local version of the many-body tensor representation (LMBTR) \cite{Huo2022}, smooth overlap of atomic positions (SOAP) \cite{Bartok2013} and atom-centered symmetry functions (ACSF) \cite{Behler2011}. For the many-body distribution functionals (MBDF) \cite{Khan2023} we used the implementation provided with the original publication. In addition, we implemented the descriptor introduced in Ref.~\citenum{Chandrasekaran2019} calling it ``Gaussian tensors'' (GT) hereafter. Finally, we used a sorted variant of the Coulomb matrix (CM) \cite{Rupp2012} similar to the bag of bonds \cite{Hansen2015} and analogous to the implementation in Ref.~\citenum{Vladyka2023}. We used the emission site S$_\mathrm{em}$ as the only center for building the descriptors LMBTR, SOAP, ACSF and GT.

We split the data set randomly to 80\% (19360 data points) for the training, and the rest 20\% (4840 data points) for testing. We calculated feature-wise z-score standardization for the obtained raw $\mathbf{D}(\mathbf{R})$ (NN input) and $\mathbf{S}$ (NN output) features using the training set and then applied this scaling to all data prior to any further procedures. The applied feature scaling is common practice in ML in general \cite{Geron2017}, and also in MD with atomic ML potentials in particular \cite{Tokita2023}. Furthermore, it can be shown that the standardization does not limit the emulation performance of an NN, but is still essential for achieving unbiased L$_2$ regularization during training (see Supplementary Information).

\subsection{Data analysis}
Successful emulation requires model selection of the hyperparameters of the NN, and also those of the descriptor family. For example, there is no single LMBTR descriptor, but the numerous internal hyperparameters of this family are subject to model selection for optimal performance. Moreover, different descriptor hyperparametrizations ({\it e.g.} those within LMBTR) require different NN hyperparameters for optimal performance of the structure--spectrum emulator system. We carried out a joint search for these two hyperparametrizations in a randomized grid search for each descriptor family separately, allocating equal computational time for each search. In each case we used the best found descriptor--NN system for all subsequent analyses. 

We used fully connected feed-forward NNs implemented in PyTorch (version 2.0.1) \cite{PyTorch} for ML. This NN architecture allows for a range of possible hyperparameters that we searched over: the weight decay term $\alpha$ (from $10^{-13}$ to $1$), number of hidden layers (2, 3, 4 or 5), hidden layer width (16, 32, 64, 128 or 256 neurons) and the learning rate (0.0001 or 0.00025). As the activation function of the neurons, we used the exponential linear unit (ELU) \cite{ELU}. The training of the networks was done in mini-batches of 200 data points by maximizing the R$^2$ score (coefficient of determination; generalized covered variance) using the Adam optimizer \cite{Adam}. We applied early stopping of the training by checking every 200 epochs if the validation score no longer improved.

We calculated the average score from a five-fold cross-validation (CV) on the training set for each selected descriptor--NN hyperparameter combination. Because building the descriptor consumed significantly more computational resources than training one NN model, we trained ten different NN architectures for every single descriptor hyperparametrization. The same procedure was repeated for each of the six descriptor families with equal total processor (CPU) time of the random grid search by allocating a total of 1440 CPU hours {\it per} descriptor family on Intel Xeon Gold 6230 processors at Puhti cluster computer, CSC, Finland. Different descriptor families have different free parameters to search over, which is detailed in the Supplementary Information. We found slight variation of the results to persist within the searched grid space due to randomness, for example, in shuffling of the mini-batches and initialization of NN weights. In the end, we chose the descriptor--NN hyperparameters with the highest CV score and trained the final model using the full training set. In this final training we used 80\% of re-shuffled training data (15488 points) for actual training and the rest 20\% (3872 points) for validation to determine the early stopping condition.

We carried out ECA decomposition \cite{Niskanen2022} of the structural descriptor space using the PyTorch implementation of the algorithm \cite{ECAclass}. In the ECA procedure, basis vectors $V=\{\mathbf{v}_j\}_{j=1}^k$ are searched for to achieve a rank-$k$ approximation for $\mathbf{D}({\mathbf R})$:
\begin{equation}
\mathbf{D}^{(k)}(\mathbf R) = \sum_{j=1}^k \underbrace{\left\langle\mathbf{v}_j|\mathbf{D}(\mathbf R)\right\rangle}_{=:t_j} \mathbf{v}_j.
\label{eq:exp}
\end{equation}
The search of the basis vectors is done by maximizing the R$^2$ score between the ML prediction for ${\mathbf{D}^{(k)}}(\mathbf{R})$ and the known data. After the basis $V$ has been established, a structure is approximated by the effective parameters $\mathbf{t}$, drastically fewer in their number than the input features. The ECA method  performs projection (\ref{eq:exp}) on the structural subspace so that the NN emulation for projected data points covers as much spectral information as possible. Noteworthily, the subspace itself (basis vectors $\mathbf{v}_j$) is optimized for the purpose, in analogy to the PCA. The procedure aims at finding structural characteristics most relevant for the spectrum outcome, and filters out degrees of freedom irrelevant in this sense. As a product, significant reduction in dimensionality of the structure--spectrum relationship is obtained; the structural information contained by 330--2700 descriptor values is reduced to less than ten $t$ scores most relevant for predicting the spectra. We optimized one basis vector at a time and applied orthonormality constraints. In the ECA procedure z-score standardization of the input features is assumed. 

\par
We found that the randomly selected initial guess of the ECA component vectors sometimes affected the resulting fit. Therefore, we ran the ECA procedure with 25 different random initial guesses for each component and always chose the vector resulting in the highest R$^2$ score, before moving on to optimizing the next component. This is likely a symptom of the high-dimensional feature vectors, which encode a wealth of atomistic information within each (correlated) element of the vector. This complexity is inherited from the physical problem itself. 
\par
We used the training set for both NN training and ECA decomposition. The generalizability of the outcome was assessed with the test set in both cases. To allow for apples-to-apples comparison, we used the R$^2$ score for all NN training, testing, and ECA. 

\section{Results}

Sulfuric acid exists in different protonation states as exemplified by sample structures in Figures~\ref{fig:spectra}a,b showing local structures from simulations of 0.9~M and 15.3~M solutions. The simulated ensemble average spectra of different concentrations are shown in Figures~\ref{fig:spectra}c,d. There is a clear concentration dependency around the K$\beta_\mathrm{x}$--K$\beta_\mathrm{1,3}$--K$\beta''$ line group, where the central peak shows a decreasing trend, and the K$\beta_\mathrm{x}$ and K$\beta''$ an increasing one, along the concentration. Additionally, our simulated spectra show shifts in energy. The general shape of the spectra match well with the experiment published in Ref.~\citenum{Niskanen2016} and, therefore, the data set can be considered suitable for the ML and subsequent analysis of this work. The chosen region and coarsened bins of the target spectra (depicted with points in Figure~\ref{fig:spectra}c,d) are sufficient to capture the shape of the spectrum.

\begin{figure*}[]
    \centering
    \includegraphics[width=\textwidth]{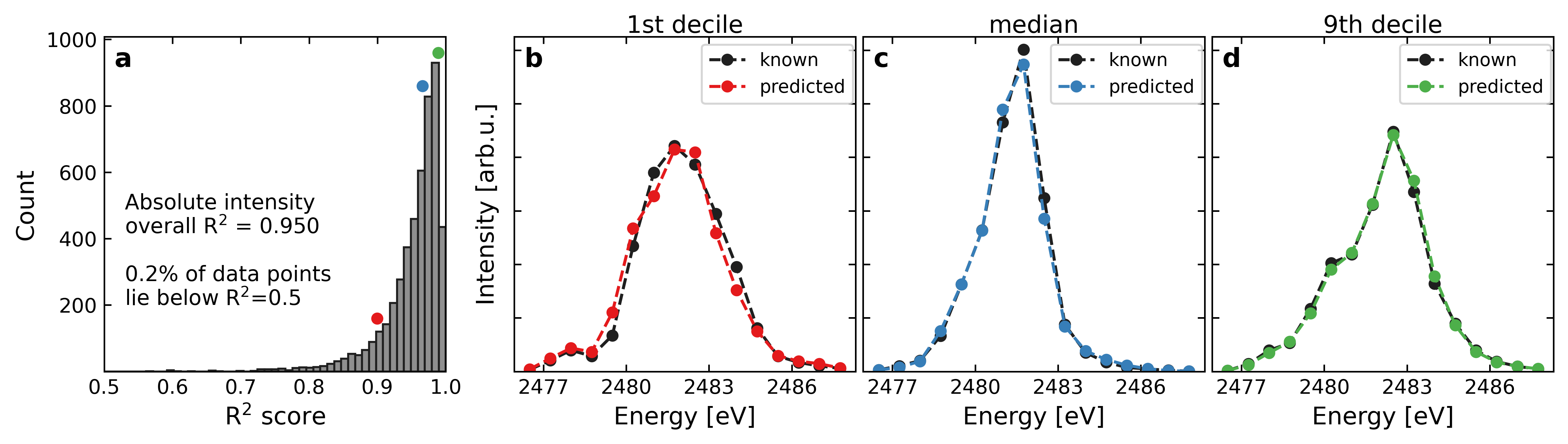}
    \caption{Absolute intensity scale spectrum prediction performance of the best NN--LMBTR model measured using the test set. {\bf a}: Distribution of R$^2$ scores for each data point. Examples of known and predicted spectra with R$^2$ score closest to {\bf b:} the 1st decile, {\bf c:} the median, and {\bf d:} the 9th decile. For each of the three cases, the location along the R$^2$ distribution is shown in panel {\bf a} as a correspondingly colored circle. The scores of this figure differ from the rest of the study as ML was applied to z-score standardized spectra. For a similar plot using the standardized spectra, see Supplementary Information.}
    \label{fig:emu_perf}
\end{figure*}

Even after extensive model selection, the descriptors yield varying prediction performance of the target spectra as presented in Table~\ref{tab:emulator_scores}. Practically equal accuracy is obtained with the best-performing descriptors LMBTR, SOAP and ACSF. The MBDF descriptor provides intermediate performance among the studied ones, whereas GT and CM yield more than 0.1 units lower R$^2$ than the most accurate descriptors. The tendency of an ML model to overfit is commonly measured by the difference between the train and the test scores. Our results hint an increasing trend in this difference along decreasing accuracy. Figure~\ref{fig:emu_perf}a illustrates the distribution of the R$^2$ scores for z-score inverse transformed (absolute intensity) spectral features using the LMBTR emulator with an overall test set R$^2$ of $0.950$. Additionally, typical prediction quality along this distribution is presented in Figure~\ref{fig:emu_perf}b--d. A similar figure for the z-score-standardized spectral space with R$^2$ of $0.928$ is available in the Supplementary Information.

\begin{table}[]
\centering
\caption{Comparison of descriptors with best performing hyperparameters after the joint model selection for the representation and neural network architecture: the number of structural features and neural network emulator train R$^2$ score, test R$^2$ score, and their difference with z-score standardized target spectra.} \label{tab:descriptor_variances}
\begin{tabular}{lcccc}\label{tab:emulator_scores}
Descriptor & N$_\mathrm{features}$ & \phantom{--} R$^2_\mathrm{train}$ \phantom{--}& R$^2_\mathrm{test}$ & Difference \\
\hline
LMBTR   & 420  & 0.944 & 0.928 & 0.015 \\
SOAP    & 2700 & 0.961 & 0.928 & 0.033 \\
ACSF    & 543  & 0.952 & 0.923 & 0.029 \\
MBDF    & 330  & 0.915 & 0.878 & 0.036 \\
GT      & 1275 & 0.857 & 0.814 & 0.043 \\
CM      & 595  & 0.889 & 0.806 & 0.083 \\
\hline
\hline
\end{tabular}
\end{table}

\begin{figure}[]
    \centering
    \includegraphics[width=\columnwidth]{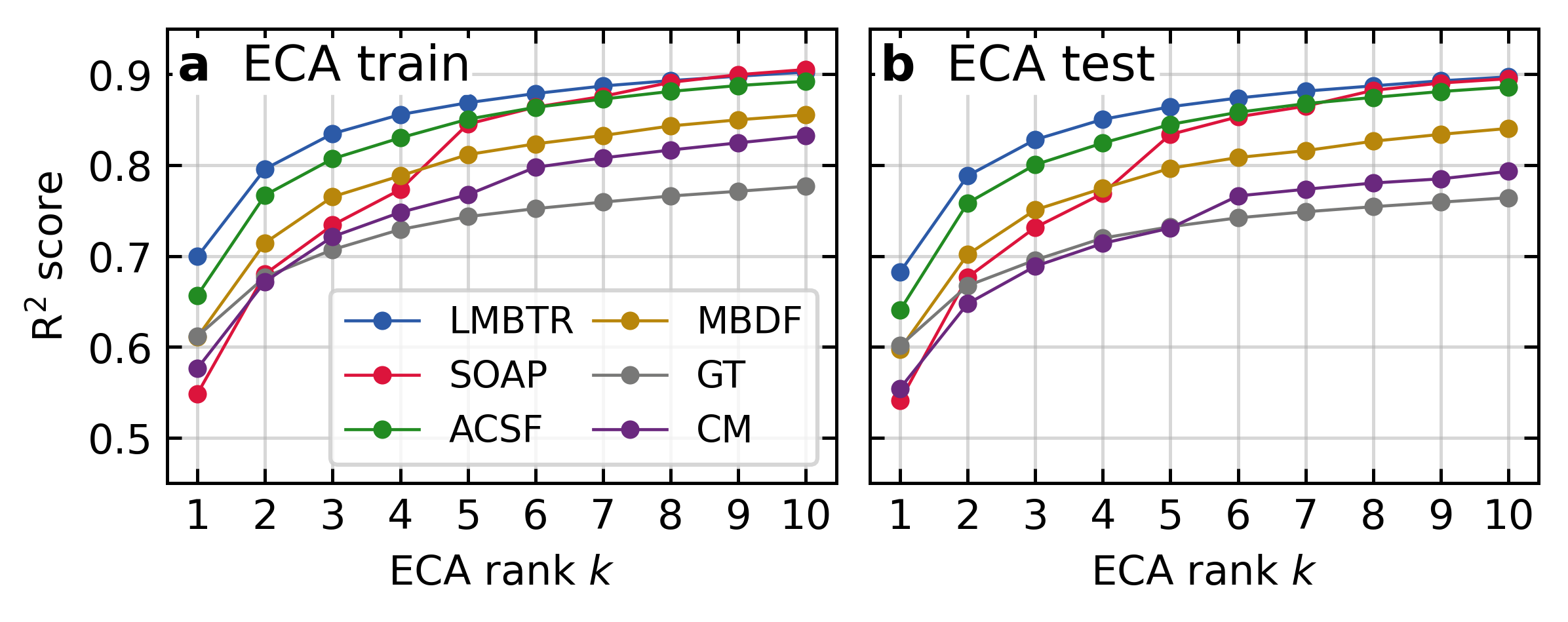}
    \caption{Typical behaviour of the R$^2$ score as a function of the ECA rank $k$ for each descriptor after joint model selection of the representation and neural network architecture. Standardized output features are used. At high ranks the results closely follow those of the emulation performance. Increasing the rank shows diminishing R$^2$-score gains. At low ranks LMBTR outperforms the rest. The jump observed SOAP and CM may occur for any descriptor and is due to the complexity of the problem and stochasticity of the iterative solution.}
    \label{fig:ECAscores}
\end{figure}

We measure the ECA decomposition performance with the R$^2$ score, shown as a function of the rank of the decomposition in Figure~\ref{fig:ECAscores}. The scores for the train set and for the test set both rise monotonically and approach a plateau near the respective emulator performance. In general, the R$^2$ scores of high rank ($\geq 5$) ECA are roughly ordered along the overall ML accuracy of the respective emulators. The design principle of ECA aims at maximal covered spectral variance at any given rank, manifested by the diminishing improvement as a function of $k$ observed in Figure~\ref{fig:ECAscores}. Consequently, the high-$k$ scores $t_k$ are not reconstructable from the spectra, as these degrees of freedom are irrelevant in their emulation. With components of negligible effect on the outcome, full structural reconstruction from spectra is impossible, as a structural descriptor is completely defined by expansion (\ref{eq:exp}) done to the full rank. The intended rapid reduction of dimensionality motivates the study of low rank ({\it e.g.} $k \le\,3$) decompositions for which LMBTR performs the best. We have noticed jumps in the R$^2$ curves as a function of the decomposition rank $k$, seen in Figure~\ref{fig:ECAscores} for SOAP and CM. This phenomenon is unpredictable and potentially related to the initial guess of the ECA component vector. Apart from the obvious complexity of the problem, the detailed origin and cure for this behaviour remain unknown to us.

\begin{figure*}[]
    \centering
    \includegraphics[width=\textwidth]{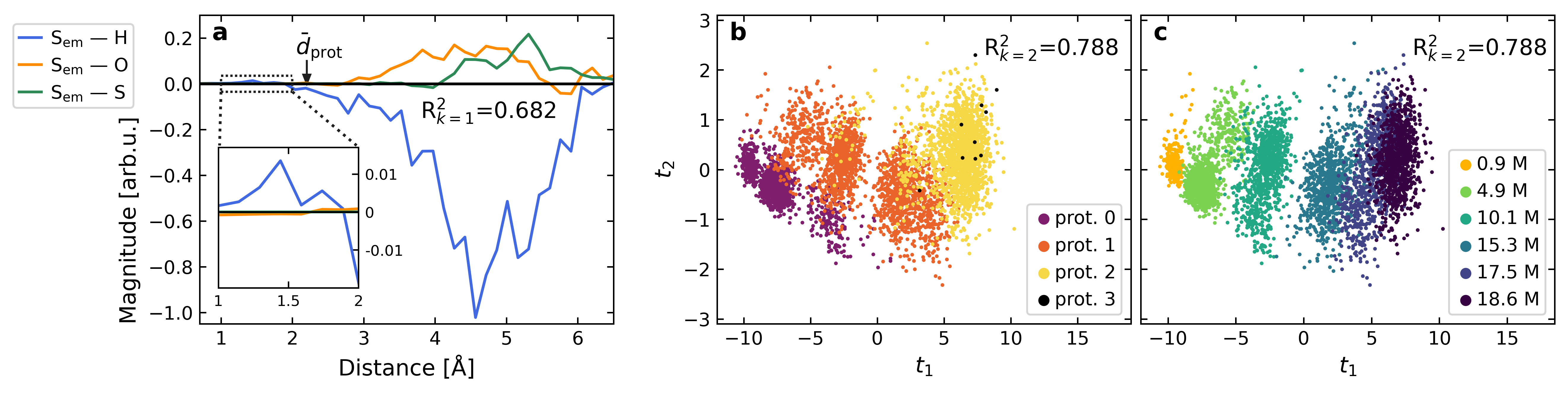}
    \caption{Results of the ECA using the LMBTR descriptor. {\bf a:} Interatomic distance part of the first ECA vector (with spectral R$^2_{k=1}=0.682$) z-score inverse transformed into the descriptor space. A separate sub-panel shows the region between 1~{\AA} and 2~{\AA}. The non-zero values of the vector show that even distant atoms have a significant effect on the spectra. The arrow at $\Bar{d}_\mathrm{prot} = 2.20$~{\AA} shows the mean distance to the hydrogen atoms protonating the acid molecule. {\bf b,c:} Two-dimensional ECA projection (with spectral R$^2_{k=2}=0.788$) of each data point in the test set: both of the components are necessary to distinguish the protonation state of the acid molecule, whereas the first component follows intermolecular interactions given by concentration. See text for details.}
    \label{fig:ECAresults}
\end{figure*}

Next, we analyse the ECA results using the LMBTR descriptor, which contains simple physical information as part of it, namely element-wise interatomic distances from the emission site S$_\mathrm{em}$. Instead of presenting the numeric values of these distances, the descriptor encodes the information on a predefined grid as a sum of Gaussian functions, centered at the respective positions. The according features of the first ECA component vector, z-score inverse transformed into the descriptor space, are shown in Figure~\ref{fig:ECAresults}a. In the sense of the aforementioned representation, these curves reflect the change in the interatomic distances most relevant in terms of the target spectrum shape. The figure shows that the S K$\beta$ XES is affected by notably distant molecules. As the typical protonating H distance from the S atom of the acid molecule is 2.2\,{\AA}, the part corresponding to the S$_{\mathrm{em}}$---H distribution shows notable relevance in the region above 3\,{\AA}. This can be attributed to the different hydrogen number density in the system reflecting the concentration, with possibly minor effects coming from hydrogen bonding of the system. The concentration dependency is further indicated by the opposite effect of S$_{\mathrm{em}}$---S curve at 4\,{\AA}--6\,{\AA}. Additionally, the S$_{\mathrm{em}}$---H distribution shows a weaker effect in the region between 1\,{\AA} and 3\,{\AA}, which likely arises from the tails of the descriptor Gaussians corresponding to the hydrogen atoms protonating the acid molecule.

The first two ECA components correspond to structural features which define the majority of the spectral variance (R$^2_{k=1}=0.682$, R$^2_{k=2}=0.788$). Recent results indicate orders of magnitude better R$^2$ score for covered spectral variance by ECA in comparison to that obtained by PCA for the structural descriptor \cite{Eronen2024}. In the case of S K$\beta$ XES of aqueous H$_2$SO$_4$ such a drastic difference is not observed: here a 2-component structural decomposition by PCA resulted in R$^2=0.453$ for spectral variance, which is still notably less than the R$^2_{k=2}=0.788$ obtained by ECA. This indicates a more direct structural characteristics -- spectral response relation in the current case.
\par
To study the separability of spectrally dominant structural features, we performed projection of each data point in the test set on two-dimensional ECA space. We focus on two most obvious characteristics: the protonation state of an acid molecule and the concentration of the system, which are indicated by coloring of each point in the resulting scatter plot, shown in Figures~\ref{fig:ECAresults}b,c, respectively. The protonation state is not the ruling structural characteristic behind variation of the S K$\beta$ XES as it can be only partially identified by the first ECA score $t_1$ (Figure~\ref{fig:ECAresults}b). However, the interplay of $t_1$ and $t_2$ disentangles these classes almost completely. This result is supported by the fact that the first component vector shows only a weak effect in Figure~\ref{fig:ECAresults}a along the part of the curve which corresponds to the protonating hydrogen atoms. In contrast, the score $t_1$ describes the concentration of the system, seen as the spectral change in Figures \ref{fig:spectra}c,d and in our analysis of the respective S$_{\mathrm{em}}$---H curve in Figure~\ref{fig:ECAresults}a. Ultimately, the need for the second degree of freedom to identify the protonation state is congruent with the overlap between the spectral-region intensity histograms of the protonation states, reported in a previous work on aqueous H$_2$SO$_4$ \cite{Niskanen2016}.

\section{Discussion}
The role of the descriptor is to present the structural data in the most utilizable form for emulation, for which numerous studies have been published in the context of atomistic systems \cite{Kwon2023,Hirai2022,Onat2020,Jager2018}. These comparisons indicate that the optimal choice of the descriptor family depends on the application, and probably on the system. For example, in prediction of the X-ray absorption near edge structure (XANES) of amorphous carbon Kwon {\it el al.} \cite{Kwon2023} found LMBTR to perform best (ACSF, LMBTR and SOAP studied). However, spectral neighbor analysis potential (SNAP) \cite{Thompson2015,Wood2018} outperformed for XANES of amorphous silicon glass in the work of Hirai {\it el al.} \cite{Hirai2022} (ACSF, LMBTR, SNAP and SOAP studied). In an other context, SOAP was deemed most favorable by Onat {\it el al.} \cite{Onat2020} for potential energy prediction of silicon (atomic cluster expansion \cite{Drautz2019}, ACSF, introduced Chebyshev polynomials in symmetry functions, MBTR, and SOAP studied). Interestingly, the same conclusion was drawn by Jäger {\it el al.} \cite{Jager2018} in prediction of hydrogen adsorption energies on nanoclusters (MBTR, SOAP and ACSF studied). We note that systematic, wide and blind hyperparameter searches appear rare in the literature. Moreover, the joint descriptor--NN hyperparameter search used in this work has further potential of improving performance of any descriptor family in such a comparison.

Our results suggest that, in X-ray spectroscopy of liquids (with $\approx 2\cdot 10^4$ data points) an equal ML performance can be obtained with LMBTR, SOAP and ACSF with joint model selection of the descriptor hyperparameters together with the NN architecture. Among the studied descriptor families, we obtained intermediate performance with MBDF, whereas we could not achieve competitive accuracy when using CM and GT. The result thus highlights the need of {\it suitable} encoding of information by the descriptor. Although the CM can be even converted back to the original structure with the loss of only handedness of the system, and although the descriptor family performed quite well for Ge K$\beta$ XES of amorphous GeO$_2$ \cite{Vladyka2023}, it does not excel with the current liquid system.

Picking a descriptor family poses a serious model selection problem because there are inherent tunable hyperparameters characteristic to each one of them (see Supplementary Information). Due to expected interplay between the optimal NN architecture, this optimization is ideally done jointly with the hyperparameters of the NN, which multiplies the required computational effort. Without prior knowledge, descriptors with more free hyperparameters are more flexible than those with fewer. Therefore, their parameter-optimized forms have a higher prior potential for accuracy as well. Because this tuning is ultimately left for the user, we accounted for the discrepancy in descriptor design by applying equal computation time for refining each descriptor, regardless how many free hyperparameters the implementation had. We propose this practice for fair assessment of structural descriptors of ever-increasing multitude. Furthermore, we find that the top-level performance among the descriptor family is typically achieved with several drastically different parametrizations. Therefore we conclude that diminishing CV score gains provides a reasonable stopping condition for the joint randomized hyperparameter search, if the allowed hyperparameter space is sufficiently large.

In this work we chose to measure the performance using the R$^2$ score, which is a widely applied metric for information captured by a model, and utilized in {\it e.g.} PCA. The score is well-suited for spectrum interpretation because it is independent of the units and of the absolute scale. The sulfur atom has a K$\beta$ baseline spectrum in the \ce{SO4} moiety, and therefore respective variation evaluated by the R$^2$ score yields an informative measure for interpretation of spectra. When using z-score standardized output, the R$^2$ gives each output feature an equal importance in the spectral interpretation, whereas raw spectral intensity favors features of large variation, observed typically for features with large overall intensity. We motivate the choice of standardization by the nonlinearity of the structure--spectrum relationship. Namely, a weak spectral feature may be indicative of a more interesting or a more widely present structural characteristics than its absolute intensity might indicate. We note however, that the analysis methods applied in this work do not necessitate the use of either output standardization or the R$^2$ score.

Analogous to PCA, ECA works as a dimensionality reduction tool. Instead of maximizing the covered structural variance, the method focuses on maximizing the spectral one for a decomposition in the structural space. As a result, the basis vectors of ECA can be used to identify descriptor features, which affect (or do not affect) the shape of the target spectra, or even for approximate structural reconstruction from spectra \cite{Vladyka2023}. The method is capable of a remarkable reduction of dimensionality \cite{Niskanen2022,Vladyka2023,Eronen2024}, improving on similar methods such as the partial least squares fitting with singular value decomposition \cite{Bookstein1996} as demonstrated in Ref.~\cite{Niskanen2022}. The first ECA component vector, shown in Figure \ref{fig:ECAresults}a, represents the dominant structural effect behind the variation of the spectrum. Although higher-rank ECA components may have cancelling contributions to those of lower ranks, these refinements are not equally relevant for spectrum interpretation as manifested by the associated diminishing spectral effect. We also note that the overall sign of an ECA vector can be chosen arbitrarily (adjusting the sign of the according score), but the relative signs of its components ({\it e.g.} the curves in Figure~\ref{fig:ECAresults}a) are always fixed.

Structural interpretation of spectra sets several requirements for a descriptor. The representation needs to allow for accurate emulation, effective decomposition, and back-conversion to simple physical information. Although all of the studied descriptors are calculated from local atomistic structures, several factors complicate recovering such information from them. These include smearing the exact values on a grid and summation of information from many atoms into one feature, possibly with distance-dependent weights. In addition, some of the descriptors rely on basis functions and may potentially have an abstract mathematical form. In this line of thought, interpretation of descriptors calls for future efforts.

Machine learning by NNs requires large data sets, that have only recently become feasible owing to the increase in computational resources and the developments in simulation tools. Advances of ML in potentials for molecular dynamics \cite{Pattnaik2020,Behler2021,Tokita2023}, and in electron structure calculations \cite{Golze2022}, could help generate more extensive and more accurate training data, leading to improved performance of spectrum emulation and subsequent analyses. 

\section{Conclusions}
We benchmarked six structural descriptor families in machine learning of simulated X-ray emission spectra (XES) of aqueous sulfuric acid. For unbiased assessment of these descriptor types with varying number of hyperparameters, we allocated equal computation time for the joint descriptor--neural network model selection in each of the six cases. We found local many-body tensor representation (LMBTR), smooth overlap of atomic positions (SOAP) and atom-centered symmetry functions (ACSF) to perform best (equally accurately) with the data set of $\sim 2\cdot 10^4$ points. 

We observed a similar hierarchy in the comparison of the descriptor families for structural dimensionality reduction guided by covered spectral variance. The LMBTR stood out especially in the low-rank decompositions of the applied emulator-based component analysis. Although the system manifests significant complexity, the analysis method managed to condense spectral dependence into two dimensions with R$^2=0.788$ for an independent test set. The results indicated that even distant atoms have a significant effect on the XES, that probes local bound orbitals around the emission site. The dominant underlying coordinate $t_1$ followed the concentration of the system, whereas inclusion of the second most relevant degree of freedom $t_2$ allowed for clear distinction of the protonation state of the acid molecule. Altogether, our results highlight loss of structural information upon formation of a spectrum, which will have implications for justified interpretation of spectra using simulations.

Structural descriptors facilitate accurate prediction of X-ray spectra by a neural network. Advances in simulation methods can be anticipated to extend and improve the data sets to allow for studies of even more complex systems and analyses with higher accuracy. Conversion of the descriptor back to simple atomistic information needs specific research efforts, as results presented in terms of these mathematically sophisticated representations can be difficult to interpret by a human.

\section*{Data availability}
The data and relevant scripts are available in Zenodo: \href{https://zenodo.org/doi/10.5281/zenodo.10650121}{10.5281/zenodo.10650121}.

\section*{Author contributions}
E.A.E. machine learning, data analysis, writing the manuscript. A.V. simulations, data analysis, writing the manuscript. C.J.S writing the manuscript. J.N. research design, simulations, funding, writing the manuscript. 

\section*{Conflicts of interest}
There are no conflicts to declare.

\begin{acknowledgments}
E.A.E. acknowledges Jenny and Antti Wihuri Foundation for funding. E.A.E., A.V. and J.N. acknowledge Academy of Finland for funding via project 331234. The authors acknowledge CSC -- IT Center for Science, Finland, and the FGCI -- Finnish Grid and Cloud Infrastructure for computational resources.
\end{acknowledgments}

\appendix

\bibliographystyle{unsrtnat}
\bibliography{references}

\newpage
\newpage
\newpage
\clearpage
\onecolumngrid
\section*{Supplementary information}
\subsection*{Spectrum simulation plane wave cutoff convergence check}
\begin{figure*}[h!!]
    \centering
    \includegraphics[width=0.4\textwidth]{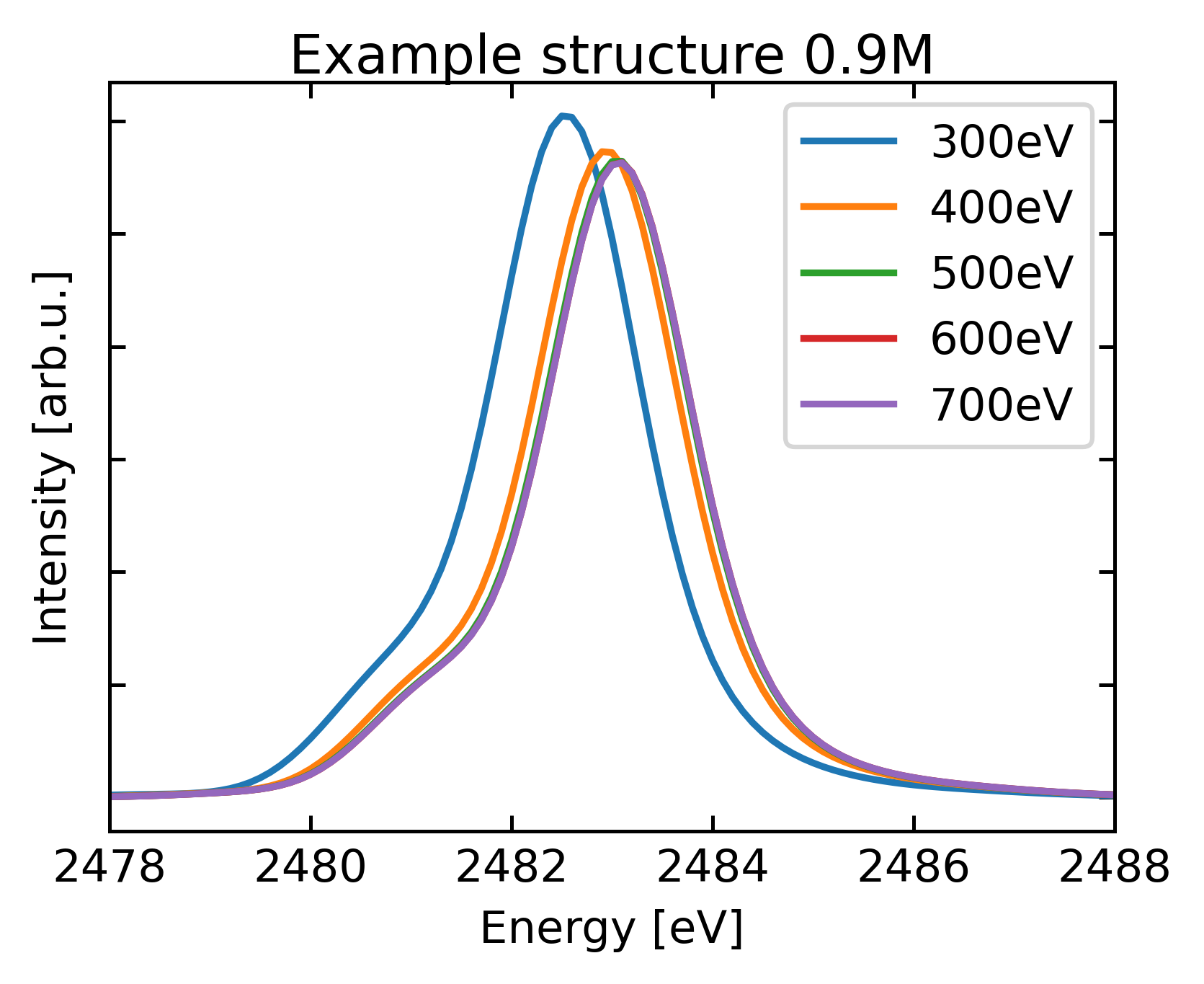}
    \caption{Spectrum from one sample structure calculated using different plane wave cutoff values. The chosen 600~eV spectrum is on top of both the 500~eV one and the 700~eV one: convergence has been reached.}
    \label{fig:PWcutoff}
\end{figure*}

We checked the convergence of the spectrum with respect to the plane wave cutoff used in the simulations. The results from one 0.9~M structure, seen in Figure~\ref{fig:PWcutoff}, show that convergence is reached after 500~eV. We used 600~eV cutoff in the simulations of this work.

\subsection*{On the role of standardization of the input}
A feedforward NN passes information from a layer (contained by input $\mathbf{x}$) to obtain activation values of the neurons of the next layer (contained by output $\mathbf{z}$) by applying function
\begin{equation}
\mathbf{z} = g(\mathbf{A}\mathbf{x}+\mathbf{b})
\end{equation}
where the weight matrix $\mathbf{A}$ and bias vector $\mathbf{b}$ are characteristic to the particular neuron layers, and are optimized during training of the network. The activation function $g$ is taken element-wise for $\mathbf{y}=\mathbf{Ax}+\mathbf{b}$
$$\begin{bmatrix}{y_1}\\\vdots\\y_{N_1}\end{bmatrix}
=
\begin{bmatrix}a_{11} & \dots & a_{1N_\mathrm{in}}\\
\vdots&\ddots&\vdots\\
a_{N_11}&\dots&a_{N_1N_\mathrm{in}}\end{bmatrix}
\begin{bmatrix}x_1\\\vdots\\x_{N_\mathrm{in}}\end{bmatrix}
+
\begin{bmatrix}b_{1} \\ \vdots \\ b_{N_1}\end{bmatrix}
.
$$
The activation of the output neuron $k$ is the activation function $z_k=g(y_k)$ of the summation
\begin{equation}
y_k = a_{k1}x_1 + \dots + a_{kN_\mathrm{in}}x_{N_\mathrm{in}} + b_{k} = \sum_{i=1}^{N_\mathrm{in}}a_{ki}x_i+ b_k.\label{activation_for_dummies}
\end{equation}
Now, say that these $\mathbf{A}$ and $\mathbf{b}$ have been found for each layer during training with non-standardized feature values. When the input features $i$ are then z-score standarized using respective mean $\mu_i$ and standard deviation $\sigma_i>0$, the input vector $\mathbf{x}'$ has components
\begin{equation}
x_i'= \frac{x_i-\mu_i}{\sigma_i}, \qquad i=1\dots N_\mathrm{in}.\label{z}
\end{equation}
In the first layer, each element $a_{ki}$ of the $k$th row of $\mathbf{A}$ have the freedom to take new value $a_{ki}' = \sigma_ia_{ki}$. Likewise, the component $b_k$ of the vector $\mathbf{b}$ has its freedom take value $b_k' = b_k+\sum_ia_{ki}\mu_i$. These adaptations for the newly formed $\mathbf{A}'$ and $\mathbf{b}'$ for the input layer yield
\begin{eqnarray}
y_k'
&=& \mathbf{A}'\mathbf{x}'+\mathbf{b}'\\
&=&
\sum_{i=1}^{N_\mathrm{in}}a_{ki}'x_i'+ b_k' \\ 
&=&\sum_{i=1}^{N_\mathrm{in}}\sigma_ia_{ki}\frac{x_i-\mu_i}{\sigma_i} + b_k+\sum_{i=1}^{N_\mathrm{in}}a_{ki}\mu_i \\
&=&\sum_{i=1}^{N_\mathrm{in}}a_{ki}x_i + b_k+\sum_{i=1}^{N_\mathrm{in}}a_{ki}\mu_i -\sum_{i=1}^{N_\mathrm{in}}a_{ki}\mu_i\\&=& \sum_{i=1}^{N_\mathrm{in}}a_{ki}x_i+ b_k = y_k
\end{eqnarray}
and the same $z_k=g(y_k)$ as without standardization. Thus the same activation of the first-layer neurons $\mathbf{z}$ (and the output of the whole NN) can always be achieved by adaptation of $\mathbf{A}$ and $\mathbf{b}$ of the input layer to the new scaling of data. In other words, transformation (\ref{z}) will not exclude any solution for training, {\it i.e.} the process of the optimization of matrices $\mathbf{A}$ and vectors $\mathbf{b}$ guided and decided by emulation performance of the NN for the training data set. We note that linear scaling of feature values $x_i$ will be a special case of Equation (\ref{z}). Moreover the result remains regardless of whether scaling is done individually for each $i$ or in coordination for groups of features $i$. This is because for each output neuron $k$ and each input feature $i$ there is a specific weight matrix element $a_{ki}$ available to adapt.
\par
However, feature scaling has an effect of the $L_2$ regularization during NN training. This regularization punishes for large $|a_{nm}|$ during the optimization of the weight matrices $\mathbf{A}$. Therefore the input feature absolute values $|x_i|$ must be kept in the same order of magnitude for all features $i$, for them all to have an equal chance to affect the summation of Equation (\ref{activation_for_dummies}). In other words, non-standardized input features would generate a systematic bias towards the information in input features larger in their numeric absolute values. We note that the z-score standardization is not only standard practice, but also a bijective mapping. Thus information is not lost in the process of z-score standardization, and it can be converted back by the respective inverse transformation. 

\subsection*{Descriptor hyperparameter grids}
The model selection for each descriptor type was carried out as a joint randomized grid search of the relevant descriptor and neural network (emulator) parameters. Every atom (S, O and H) of any molecule within $6~\text{\AA}$ of the emission site was included in constructing the descriptor vector, except in the case of CM. A single center, the emission site S$_\mathrm{em}$, was used with the descriptors LMBTR, SOAP, ACSF and GT. When applicable, the cut of radius "r cut" was selected as $6.5~\text{\AA}$ to include all the atoms which were part of the spectrum calculations.

\subsubsection*{LMBTR}
We used the implementation of the DScribe package \cite{Himanen2020,dscribe2} for the LMBTR descriptor \cite{Huo2022}.  The studied parameter grid is presented in Table \ref{tab:lmbtr}. Our previous knowledge of the descriptor \cite{Eronen2024} helped with the selection of the grid. Some unnecessary features (always zero as our system is non-periodic) were removed from the output.
\begin{table}[h!!]
\caption{The LMBTR parameter grid. Where applicable, the selected parameter value is shown in bold.}
\begin{tabular}{c | c | c} 
Parameter & Distances (k2) & Angles (k3)  \\ \hline
geometry function & distance & angle \\
grid min & $0.7~\text{\AA}$ & $0^\circ$\\
grid max & $6.5~\text{\AA}$ & $180^\circ$\\
grid n & 20,{\bf 40},60,80,100 & 5,10,{\bf 20},30,40\\
grid $\sigma$ & 0.2,0.4,{\bf 0.6},0.8,1.0$~\text{\AA}$ & 9,12,{\bf 15},18,21$^\circ$\\
weighting function & unity & exp\\
weighting scale & --- & 0.8,1.0,1.2,{\bf 1.4},1.6\\
weighting threshold & --- & 1e--8\\
\end{tabular}
\label{tab:lmbtr}
\end{table}

\subsubsection*{SOAP}
We used the implementation of the DScribe package \cite{Himanen2020,dscribe2} for the SOAP \cite{Bartok2013} descriptor. The studied parameter grid is presented in Table \ref{tab:soap}.
\begin{table}[h!!]
\caption{The SOAP parameter grid. Where applicable, the selected parameter value is shown in bold.}
\begin{tabular}{c | c } 
Parameter & Values   \\ \hline
r cut & $6.5~\text{\AA}$\\
n max & 4,5,6,7,{\bf 8}\\
l max & 4,5,6,7,{\bf 8}\\
sigma & 0.25,{\bf 0.5},0.75,1.0$~\text{\AA}$\\
rbf & gto\\
weighting & pow\\
c & {\bf 0.25},0.5,1.0,2.0,4.0\\
d & 0.25,0.5,{\bf 1.0},2.0,4.0\\
m & {\bf 2},4,6,8\\
r0 & 1,2,3,{\bf 4}\\
\end{tabular}
\label{tab:soap}
\end{table}

\subsubsection*{ACSF}
We used the implementation of the DScribe package \cite{Himanen2020,dscribe2} for the ACSF \cite{Behler2011} descriptor. The studied parameter grid, inspired by Nguyen and co-workers \cite{Nguyen2018}, is presented in Table \ref{tab:acsf}. We used $G^2$ and $G^4$ to include both two-body and three-body (radial and angular) interactions. For $G^2$ we used "$R_s$ n" linearly spaced values from $0.7~\text{\AA}$ to $6.5~\text{\AA}$ and $\eta = 12.5/(R_s^2)$. For $G^4$ we always had both $\lambda = -1 $ and $\lambda = 1$, and $\zeta = 2^x$ where $x$ is a range of integers from $0$ to some value with increments of $1$. The parameter $\eta$ had "$\eta$ n" values from "$\eta$ min" to "$\eta$ max" placed on a logarithmic grid.

\begin{table}[h!!]
\caption{The ACSF parameter grid. Where applicable, the selected parameter value is shown in bold. "$R_s$ n" is for the $G^2$ function and the rest are for the $G^4$.}
\begin{tabular}{c | c } 
Parameter & Values   \\ \hline
 r cut & $6.5~\text{\AA}$\\
$R_s$ n  & 10,20,30,40,50,{\bf 60},70,80 \\
$\eta$ min & 0.0001,0.001,{\bf 0.01},0.1 \\
$\eta$ max & {\bf 1.0},2.0,3.0,4.0 \\
$\eta$ n & {\bf 5},10,15,20,25,30\\
$\zeta$ min & 1\\
$\zeta$ max & 32,{\bf 64},128,256\\
$\lambda$ &$-1$~and~$1$ \\
\end{tabular}
\label{tab:acsf}
\end{table}
\subsubsection*{GT}
The parameter grid for the self-implemented descriptor GT \cite{Chandrasekaran2019} is presented in Table \ref{tab:gt}. Each of the three components (scalar, vector and tensor) had the same minimum and maximum gaussian width values. The number of linearly spaced widths between the minimum and the maximum ("n") was unique for each component.

\begin{table}[h!!]
\caption{The GT parameter grid. Where applicable, the selected parameter value is shown in bold.}
\begin{tabular}{c | c } 
Parameter & Values   \\ \hline
 r cut & $6.5~\text{\AA}$\\
width min & 0.025,0.05,0.1,0.2,{\bf 0.4}$~\text{\AA}$ \\
width max & 4,5,6,{\bf 7},8,9,10$~\text{\AA}$ \\
scalar n & 10,25,50,75,100,{\bf 125},150\\
vector n & 10,20,40,{\bf 60},80,100 \\
tensor n & 10,20,40,60,{\bf 80},100 \\
\end{tabular}
\label{tab:gt}
\end{table}

\subsubsection*{CM}
The parameter grid for the self-implemented descriptor CM \cite{Rupp2012,Hansen2015} is presented in Table \ref{tab:cm}. In our implementation the columns and rows of the initial full Coulomb matrix are first arranged by grouping the elements. These groups are then individually distance sorted with respect to the emission site. For each element S, O and H, only a specific number of the closest atoms (with respect to the emission site $S_\mathrm{em}$) are included in the final matrix. If a given structure had less atoms of a given element than specified in the grid, the corresponding matrix elements were zeroed. We constructed the used descriptor vector by flattening the upper triangle (excluding the diagonal) of the Coulomb matrix.

\begin{table}[h!!]
\caption{The CM parameter grid. The selected parameter value is shown in bold.}
\begin{tabular}{c | c } 
Parameter & Values   \\ \hline
sulfur n  & 1,2,3,4,5,6,{\bf 7},8,9,10,11,12,13,14,15 \\
oxygen n   & 4,6,8,10,12,{\bf 14},16,18,20,22,24,26,28,30,32,34,36,38,40,42,44,46,48,50 \\
hydrogen n & 2,6,10,{\bf 14},18,22,26,30,34,38,42,46,50,54,58,62,66\\
\end{tabular}
\label{tab:cm}
\end{table}

\subsubsection*{MBDF}
The parameter grid for the MBDF developed and implemented by Khan {\it et al.} \cite{Khan2023} is presented in Table~\ref{tab:mbdf}. Excluding the cutoff radius "r cut", the internal parameters of the descriptor selected by the original authors were assumed to be suitable. Ideally, we would have included these in the search, but this would have made the model selection computationally too heavy as building the feature vector required significantly more CPU time than the other descriptors. Instead, we opted for an approach similar to that applied to CM. For each atom (row), the initial matrix had a total of six features (columns). The rows of the matrix are first arranged by grouping the elements. These groups are then individually distance sorted with respect to the emission site S$_\mathrm{em}$. For each element S, O and H, only a specific number rows (obtained from the closest atoms with respect to the emission site) were included in the final matrix. In the end, all the atoms (within $6~\text{\AA}$ of the emission site) are included in the descriptor through at least one row, as each row contains information about the local environment within "r cut" of one specific center atom. If a given structure had atoms of a given element than specified in the grid, the corresponding matrix elements were zeroed.

The authors also introduced a grid-based variant, which is independent of the number of atoms. The variant caused an increase in the dimensionality of the feature vector including a notable portion of zeros, and was left out of this work. At the time of download (2023-11-08), the provided code produced six features for each included center atom.

\begin{table}[h!!]
\caption{The MBDF parameter grid. Where applicable, the selected parameter value is shown in bold.}
\begin{tabular}{c | c } 
Parameter & Values   \\ \hline
 r cut & $6.5~\text{\AA}$\\
sulfur n  & 1,2,3,4,5,6,7,8,9,10,11,12,{\bf 13},14,15 \\
oxygen n   & 4,6,8,10,12,14,16,18,20,22,24,26,28,30,32,34,{\bf 36},38,40,42,44,46,48,50 \\
hydrogen n & 2,{\bf 6},10,14,18,22,26,30,34,38,42,46,50,54,58,62,66\\
\end{tabular}
\label{tab:mbdf}
\end{table}

\newpage
\subsection*{Hyperparameter grid search R$^2$-score distributions}
\begin{figure*}[h!!]
    \centering
    \includegraphics[width=\textwidth]{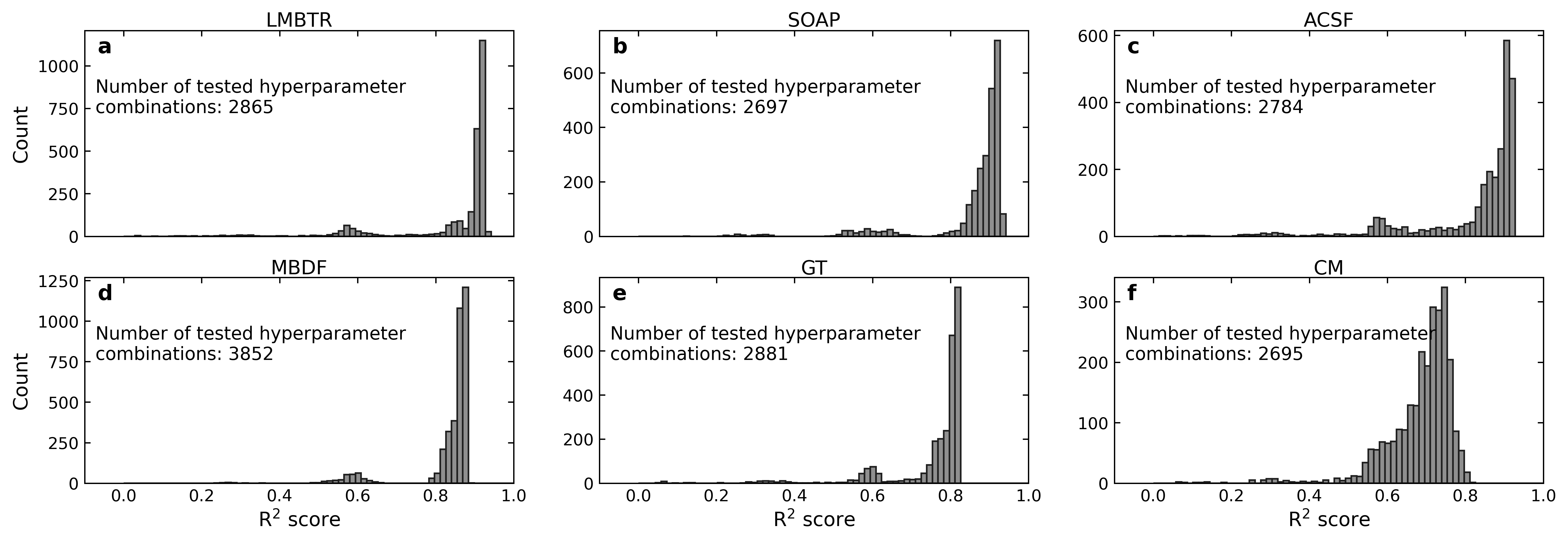}
    \caption{{\bf a--f:} The CV R$^2$-score distributions and the number of tested hyperparameter combinations for each of the six descriptors. We used the same bins for each panel.}
    \label{fig:MS}
\end{figure*}

The cross validation R$^2$-score distributions and the number of tested hyperparameter combinations for each of the six descriptors is shown in Figure~\ref{fig:MS}.

\subsection*{Visualization of the performance of the best model with standardized spectra }
\begin{figure*}[h!!]
    \centering
    \includegraphics[width=\textwidth]{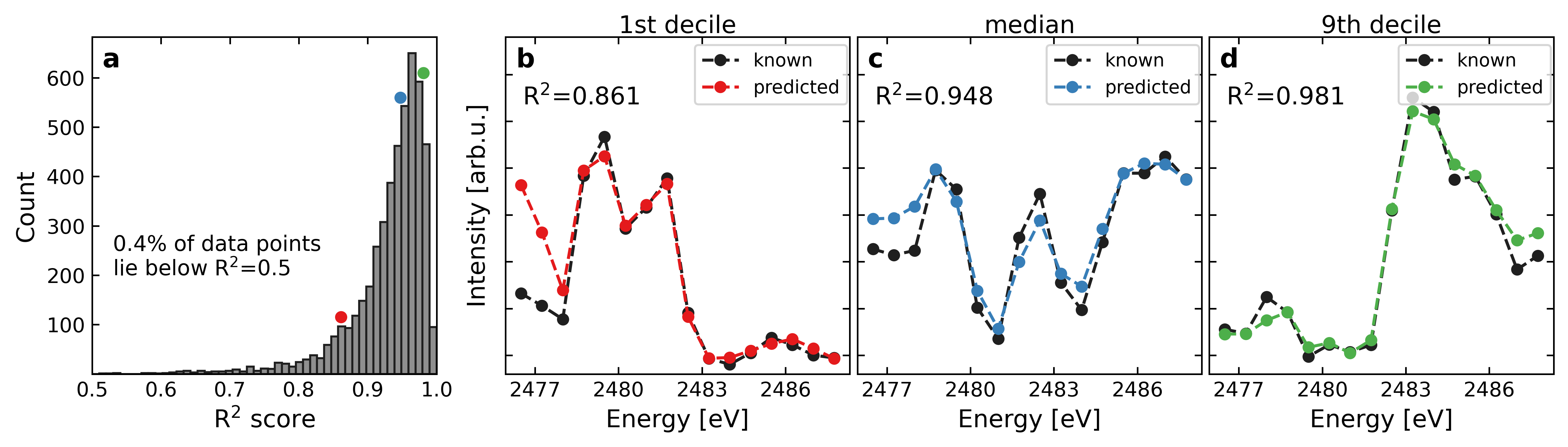}
    \caption{Performance of the best NN--LMBTR on the test set with z-score standardized spectra. {\bf a}: Distribution of R$^2$ scores for each data point. Examples of known and predicted spectra with R$^2$ score closest to {\bf b:} the 1st decile, {\bf c:} the median, and {\bf d:} the 9th decile. For each of the three cases, the location along the R$^2$ distribution is shown in panel {\bf a} as a correspondingly colored circle.}
    \label{fig:spec_suppl}
\end{figure*}

Visualization of the performance of the best NN--LMBTR model on the z-score standardized test set spectra similar to Figure~2 of the main text is shown in Figure~\ref{fig:spec_suppl}.

\end{document}